\documentclass{JAC2003}
\usepackage{graphicx}
\usepackage{booktabs}

\begin{document}
\title{GRID INDUCED NOISE AND ENTROPY GROWTH IN 3D PARTICLE-IN-CELL SIMULATION OF HIGH INTENSITY BEAMS\\[-.7\baselineskip]}

\author{I. Hofmann and O. Boine-Frankenheim \\  GSI Helmholtzzentrum f\"{u}r Schwerionenforschung GmbH, Planckstr. 1,  64291 Darmstadt, Germany \\
Technische Universit\"{a}t Darmstadt, Schlossgartenstrasse 8, 64289 Darmstadt, Germany}
 
\maketitle

\begin{abstract}
The numerical noise inherent to particle-in-cell simulation of 3D high intensity bunched beams is studied with the TRACEWIN code and compared with the analytical model by Struckmeier~\cite{struck1994}. The latter assumes that entropy growth can be related to Markov type stochastic processes due to temperature anisotropy and the artificial "collisions" caused by using macro-particles and calculating the space charge effect. The resulting noise can lead to growth of the six-dimensional rms emittance and a suitably defined rms entropy. We confirm the dependence of this growth on the bunch temperature anisotropy as predicted by Struckmeier. However, we also find an apparently modified mechanism of noise generation by the non-Liouvillean effect of the Poisson solver grid, which exists in periodic focusing systems even in the complete absence of temperature anisotropy.  Our findings are applicable in particular to high current linac simulation, where they can provide guidance to estimate noise effects and help finding an effective balance between the number of simulation particles and the grid resolution.
\end{abstract}

\section{Introduction}
In modelling of high intensity beams by particle-in-cell (PIC) computer simulation it is of importance to understand  whether any observed emittance growth or beam loss is caused by real physical processes,  or by non-physical processes like numerical discretisation and grid effects from the Poisson solver, which can be also seen as pseudo-collisions with the grid charge distribution. Our emphasis is on the numerical noise generated by the discreteness of the spatial grid used for the Poisson solver and the finite number of particles. We focus on beam parameters typical for high intensity linear accelerators, but our findings can also be extended to circular accelerators with very different ratio of transverse to longitudinal parameters. 

It is commonly accepted that real physical processes leading to amplitude and emittance growth can be initial mismatch; mismatch at structure transitions; errors; external or internal resonances; but also the fact that with space charge strictly periodic stationary distribution functions are not explicitly known in 3d (for a detailed discussion see Ref.~\cite{struck1992}). In storage rings intra-beam scattering is often observed as another kind of real physical process, which leads to a total emittance growth under certain circumstances. Numerical noise, on the other hand, can have a  similar effect on the beam as real intra-beam scattering. It is one of the challenges of high intensity beam simulation to be able to distinguish between physical and numerical growth effects. Towards this end an in-depth understanding and parametrisation of numerical noise is crucial. 

Collision or noise effects can, in principle, be associated with entropy growth. The concept of entropy in accelerator beams was first discussed in general terms by Lawson et al.~\cite{lawson1973}. Later on the Fokker-Planck equation - see Ref.~\cite{jansen1990} for a detailed discussion of its applicability to intra-beam scattering - was employed to analyse mismatch and turbulent heating of a sheet beam by Bohn~\cite{bohn1993}. A significant step ahead in this direction has been the noise and entropy growth model by Struckmeier~\cite{struck1996,struck2000}, which provides a useful theoretical framework based on second order moments of the Vlasov-Fokker-Planck equation. It assumes collisional behaviour and  temperature anisotropy to drive 6d rms emittance growth, which is used to explain rms entropy growth. 

Evidence of such an emittance growth by grid and particle number related "spurious collisions" in a linac environment - using the PICNIC space charge routine within the PARMILA code - was given in Ref.~\cite{pichoff1998}. The present computational study makes use of the analytical framework by Struckmeier, examines its validity for 3d bunched beams under space charge conditions typical for high intensity proton or ion linear accelerators, and presents different parametric dependencies to serve as basis for optimization. For our study  we use the TRACEWIN PIC  code~\cite{uriot} developed primarily for linear accelerators. It also employs the PICNIC space charge routine with an $rz$ as well as an $ xyz $  option. Detailed comparison with results from other codes or simulation codes more suitable for circular accelerators is left to future work.

\section{Rms approach to entropy growth}\label{sec:rms-approach}
It is worth reminding the fact that in analytical  descriptions of high intensity beam  phenomena it is generally assumed that the flow of particles in 6d phase space can be described by a Hamiltonian with - in general - time-dependent external forces and a time-dependent self-consistent space charge potential. Liouville's theorem applies, which implies that the volume of a phase space element occupied by particles remains invariant in time. In such a system there is no entropy growth, if infinite phase space resolution is assumed and the Boltzmann entropy as logarithm of probabilities in 6d or $\mu$ space is evaluated accordingly. 

The same applies to a Hamiltonian flow in the 6N-dimensional phase space ($\Gamma$ space), which includes systems with Coulomb collisions. Growth of the Gibbs entropy (logarithm of probabilities in $\Gamma$ space) requires introduction of a "coarse-grained" phase space as opposed to the "fine-grained" infinite resolution phase space, which of course breaks the Hamiltonian nature of the flow. There is a certain analogy between coarse-graining in $\Gamma$ space and the PIC-code technique of distributing particles on a 3d grid to solve Poisson's equation. The latter also breaks the Hamiltonian nature of the flow in 6d phase space and is a source of entropy growth. Under certain circumstances the interaction with this grid can be also looked at  as "collision" with the grid charge distribution.

The idea of a probability based approach to entropy  by using the logarithm of the rms emittance was introduced by Lawson et al.~\cite{lawson1973}. For a time-independent Kapchinskij-Vladimirskij distribution in 4d phase space they thus obtained
\begin{equation}\label{lawson}
S=k \:\ln\:\epsilon,
\end{equation}
where $k$ is the Boltzmann constant.

An  important step towards dynamically evolving distributions has been the approach by Struckmeier~\cite{struck1996,struck2000}. He demonstrated that the rms - second order moments - approach originally introduced by Sacherer~\cite{sacherer1971} to derive envelope equations, and later generalized to include the field energy in 6d (see Ref.~\cite{hofmann1987}), can be extended to the full Vlasov-Fokker-Planck equation. The thus obtained equation applies to the product of the three rms emittances, which can be understood as a six-dimensional rms emittance, $\epsilon_{6d}\equiv \epsilon_{x} \epsilon_{y} \epsilon_{z} $, hence
\begin{eqnarray}\label{emit-evol}
\nonumber
\frac{d}{ds}\ln\:\epsilon_{x}(s)\epsilon_{y}(s)\epsilon_{z}(s)=\frac{k_{f}}{3}\:I_A=\\
\frac{k_{f}}{3}\left( \frac{(1-r_{xy})^{2}}{r_{xy}}+\frac{(1-r_{xz})^{2}}{r_{xz}}+\frac{(1-r_{yz})^{2}}{r_{yz}}\right)\geq 0 
\end{eqnarray}
where $ s=\beta ct $ measures the distance and $k_{f}\equiv \beta_{f}/\beta c\gamma$, with $\beta_{f}$ the dynamical friction coefficient, which is proportional to the Coulomb logarithm. The $ r_{nm} $ are rms based "temperature ratios" or ratios of intrinsic spreads of velocities given here in non-relativistic approximation:
\begin{equation}\label{anisotropies}
r_{xy}(s)\equiv \frac{T_y(s)}{T_x(s)}, \:r_{xz}(s)\equiv \frac{T_z(s)}{T_x(s)}, \:r_{yz}(s)\equiv \frac{T_z(s)}{T_y(s)},
\end{equation}
For upright ellipses we simply have the familiar rms expressions
\begin{equation} \label{ratio}
\frac{T_y(s)}{T_x(s)}\equiv \frac{\epsilon_{y}^{2}/\left\langle y^2\right\rangle   }{\epsilon_{x}^{2}/\left\langle x^2\right\rangle }=\frac{\epsilon_{y}k_y}{\epsilon_{x}k_x}
\end{equation}
and similar for the remaining ratios.
Some caution is necessary when using a concept like "temperature". Strictly speaking, it cannot be defined properly for our beams, which are not in an equilibrium state. In the above model the concept of locally near-isotropic temperatures is formally adopted as basis for the validity of the Einstein relation connecting isotropic diffusion and friction coefficients, which is only valid for not too large deviation from thermal equilibrium. 

It is important to note that Eq.~\ref{emit-evol} suggests a separation of a "friction term" (given by $ k_{f} $) and a "temperature anisotropy term" $I_A$ defined by the bracket on the r.h.s. of Eq.~\ref{emit-evol}. For the present study we shall not be concerned further about a more rigorous justification of $ k_{f} $, which is left to a forthcoming study. Instead, we adopt a phenomenological approach with the goal to  demonstrate that this separation is a practical ansatz. The details of the numerical scheme and of the Coulomb logarithm are assumed to enter into the coefficient $ k_{f} $), which is separated from the "driving" anisotropy term. Note that - as one may expect from extended plasmas - no growth of the rms emittance is predicted from  Eq.~\ref{emit-evol}, if all temperatures are identical everywhere and the r.h.s. vanishes. This issue will be examined and discussed further below.

As the r.h.s. of Eq.~\ref{emit-evol} is either zero or positive, $\ln \epsilon_{x}\epsilon_{y}\epsilon_{z}$ cannot decrease. Noting that the probability of decoupled events is the product of individual probabilities and that entropy should be an "extensive" (i.e. additive) quantity, it is justified to relate this logarithm to the entropy in an rms sense analogous to the static KV-case in Ref.~\cite{lawson1973}:
\begin{eqnarray}\label{entropy}
\frac{1}{k}\frac{dS}{ds}=\frac{d}{ds}\ln\:\epsilon_{x}(s)\epsilon_{y}(s)\epsilon_{z}(s)
=\frac{k_f}{3}\:I_A\geq 0 
\end{eqnarray}
In the remainder of this work growth of $ \epsilon_{6d} $ is thus used as a synonym to entropy growth. It is, however, necessary to apply some caution here. An entropy definition based on rms quantities cannot be applied to collective or resonant processes beyond second order, which may also cause growth of rms emittances, but at the same time a decrease of $ \epsilon_{6d}$. This will be discussed in more detail in the following section.

\section{3d bunches in periodic solenoid with $rz$ Poisson solver}
\label{sec:numerical-rz}
We employ the TRACEWIN code~\cite{uriot} with a bunched
beam, and  in all examples of this section a periodic lattice with thin solenoid lenses and thin rf
gaps  at the same location. Due to the rotational symmetry the $rz$ Poisson solver option of TRACEWIN/PICNIC is sufficient.  It is characterized by the number $n_{cr} $ of radial cells from the origin to $R_{max}$; and $n_{cz} $ as radial cells from $-Z_{max}$ to $Z_{max}$. For these maximum grid extent values we have assumed throughout the values 3$\sigma$, where $\sigma$ is the rms size in each direction. Beyond the grid limits PICNIC determines fields analytically assuming a Gaussian density distribution of the same rms sizes. 

\begin{figure}[h]
   \centering
   \includegraphics*[width=85mm]{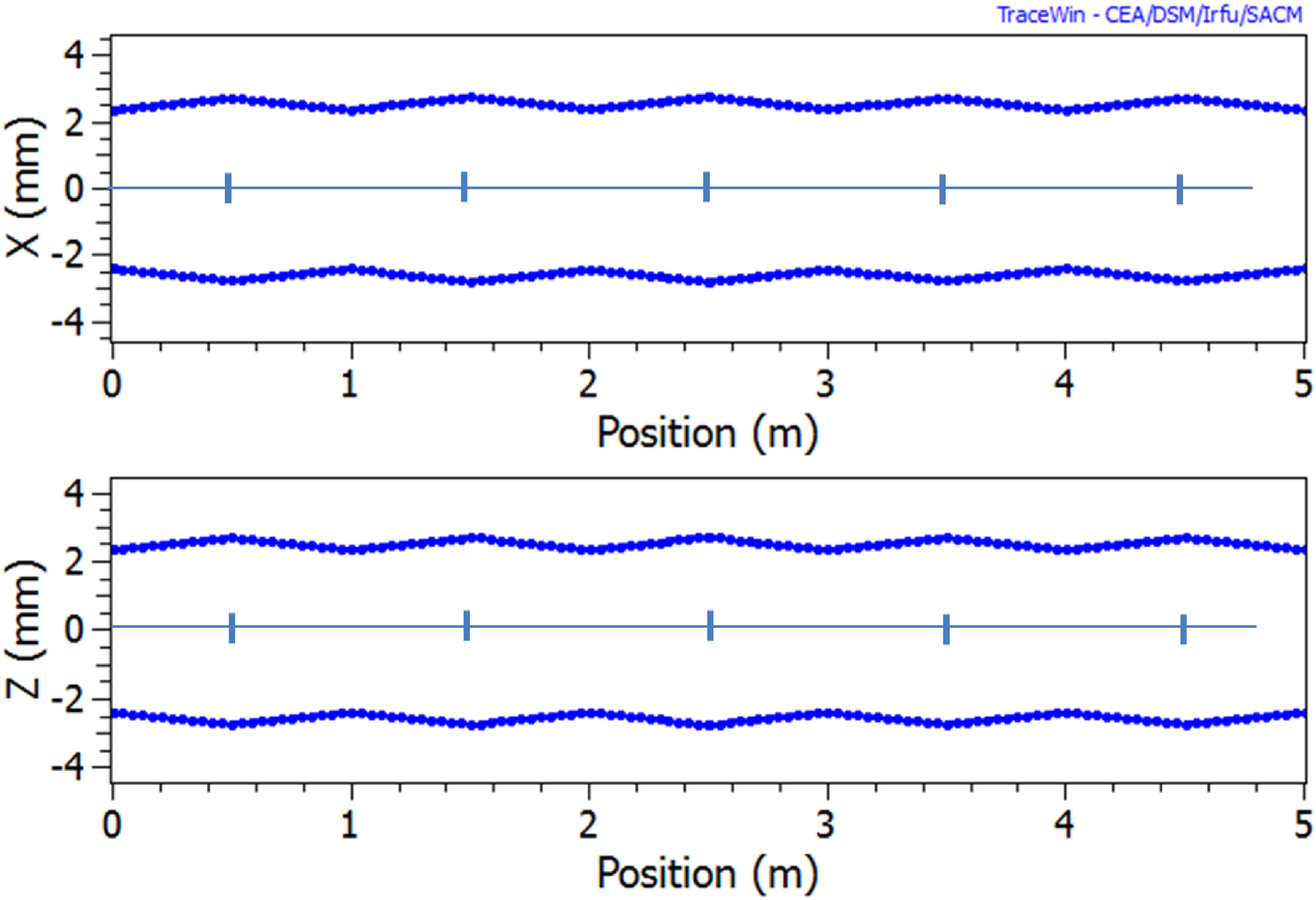}
   \caption{Envelope of spherical bunch in standard periodic solenoid lattice with bars indicating positions of combined solenoid and rf kicks (5 periodic cells).}
   \label{env}
\end{figure}

Typical matched periodic envelopes over 5 cells - assuming a length of 1 m per cell - for a spherical (breathing) bunch with equal emittances in $x,y,z$ and $k_{0x,y,z}=60^{o}$ are shown in Fig.~\ref{env}. In all cases - unless mentioned otherwise - we set the option "steps of calculation" to 20/10 in TRACEWIN, which implies 20 lattice calculations and 10 space charge steps per meter of drift space, in addition to one lattice calculation and one space charge step per solenoid. Due to a rounding up algorithm this amounts to 14 space charge steps per lattice cell, which is found as sufficient. As starting distribution function we employ either the TRACEWIN standard option "6d-ellipsoid" (randomly generated in the six-dimensional phase space ellipsoid with parabolic density profiles) or the option "Gaussian" (Parmila type 36).

\subsection{Distinction of "coherent" and "noise dominated" regimes}
\label{sec:distinction}
\begin{figure}[h]
   \centering
   \includegraphics*[width=75mm]{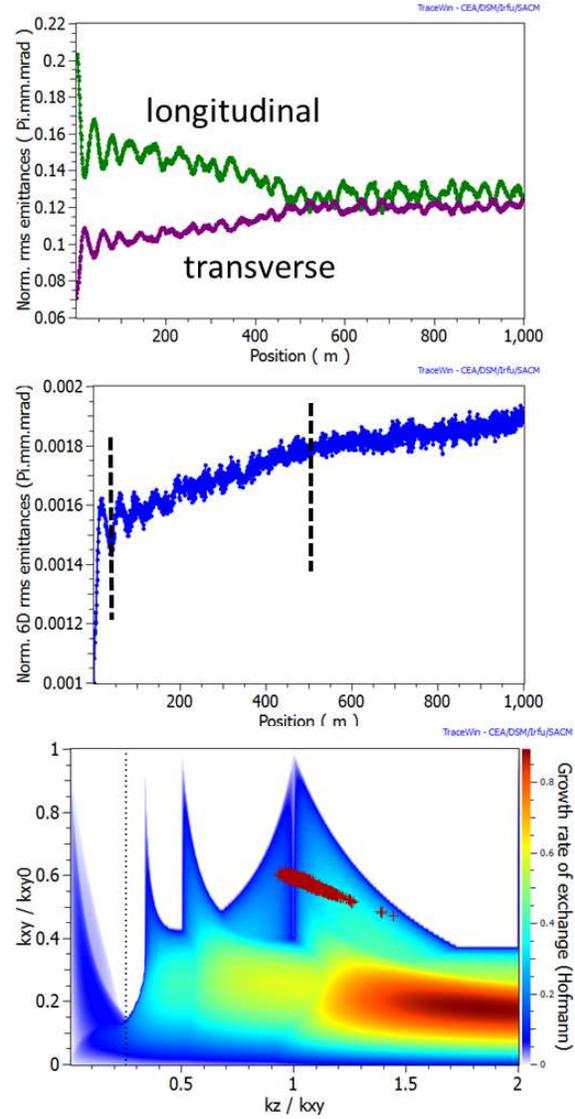}
   \caption{Rms emittances (top); $ \epsilon_{6d} $ (center, with dashed lines indicating transitions between different regimes); and stability chart (bottom) for $ \epsilon_{z}/\epsilon_{xy}=3$ including tune footprint over 1000 periodic cells (m).}
   \label{regimes}
\end{figure}

As mentioned above
the validity of the rms entropy evolution is limited to incoherent noise effects. A distinction between essentially three regimes is recognized in Fig.~\ref{regimes}, where the noise effect is enhanced by choosing a particle number as low as 1000. An initial anisotropy with
 $\epsilon_{x,y}/\epsilon_{z}=0.35$ is assumed for a zero current phase advance $k_{0x,y,z}=60^{o}$, which results in space charge depressed tunes of $k_{x,y,z}=32/32/38^{o}$.  The starting distribution function is the "6d-ellipsoid" option. Results of the emittance evolution are shown in Fig.~\ref{regimes}.
Three regimes of emittance behaviour are identified: 
\begin{enumerate}
\item \textbf{Coherent regime:} It is noted that in less than 20 cells a partial exchange  occurs between transverse and longitudinal emittances. This is caused by the coherent space charge driven fourth order resonance with a stop-band around $ k_z/k_{x,y} =1$ as shown in the stability chart of Fig.~\ref{regimes} (details see Ref.~\cite{hofmann2003}). The selfconsistently calculated tune foot print (evolving from larger to smaller values of $ k_z/k_{x,y}$ due to the dynamically changing emittances) indicates the rapid initial changes of the tune ratio over the first few cells. During this resonant phase there is a steep increase of $ \epsilon_{6d} $, followed by pronounced oscillations, and $ \epsilon_{6d} $ cannot be interpreted as entropy. This is consistent with the observation that the underlying dynamics takes place beyond second order. Note that the only partial approach to equipartition in this regime is related to the starting tune ratio at the right edge of the stop-band in Fig.~\ref{regimes}.
\item \textbf{Anisotropy induced noise:} The remaining anisotropy slowly vanishes in the subsequent \textit{noise and  anisotropy} driven regime, which gradually leads to equal emittances and tunes at about 500 m, where $k_{x,y,z}\approx 36^{o}$. Hence the   beam becomes fully equipartitioned, and the vanishing r.h.s. of Eq.~\ref{emit-evol} predicts no further emittance or entropy growth. 
\item \textbf{Grid induced noise:} The continuing growth of $ \epsilon_{6d} $ beyond the point of equipartition is unexpected in the frame of Eq.~\ref{emit-evol}. We assume this growth is caused by a violation of the assumption of a locally isotropic collision type diffusion process - the underlying assumption in the derivation of Eq.~\ref{emit-evol} - in our periodic focussing system. It must be assumed that this violation is actually a combined effect of several mechanisms:
\begin{itemize}

\item Coulomb interaction is via charges deposited on the grid rather than direct particle-particle interaction; the distribution of charges on a grid introduces a "coarse-graining" component, which is not subject to Hamiltonian mechanics, hence Liouville doesn't apply and entropy may grow.
\item The periodic modulation of focusing introduces a "coherent" streaming against the grid - even in a fully isotropic situation.
\item The modulation of focussing is not slow compared with "collision times".  
\end{itemize}
This suggests that it may be appropriate to introduce a purely grid-related noise term $I_{GN}$ in Eq.~\ref{entropy}, which is independent of the temperature anisotropy term: 
\begin{eqnarray}\label{modified}
\frac{d}{ds}\ln\:\epsilon_{x}(s)\epsilon_{y}(s)\epsilon_{z}(s)
=\frac{k^{^\star}_f}{3}\:\left(I_A+I_{GN}\right).
\end{eqnarray}
Note that we have introduced a $ k^{^\star}_f $ to take into account that these grid related effects might also have an effect on $ k_f $, besides the offset term $ I_{GN} $. 

Derivation of such a modified equation from first principles is beyond the scope of this paper. Instead, we take this phenomenological approach and present further support to it in the following sections. 
\end{enumerate}

\subsection{Scaling with intensity}\label{sec:intensity}
\begin{figure}[h]
   \centering
   \includegraphics*[width=85mm]{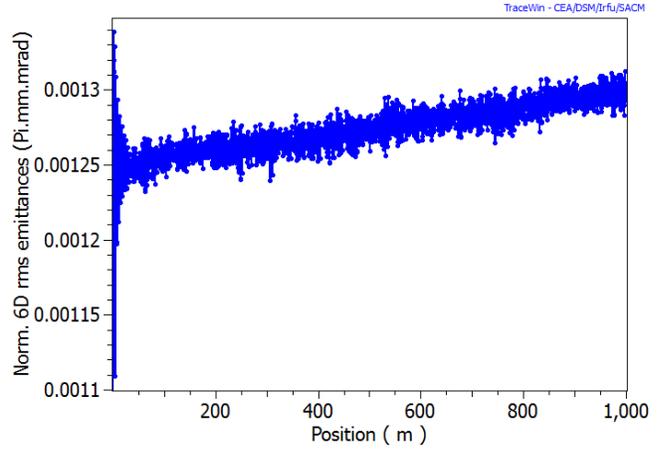}
   \caption{Example of evolution of $\epsilon_{6d}$ for N=4000.}
   \label{ramp}
\end{figure}

\begin{figure}[h]
   \centering
   \includegraphics*[width=85mm]{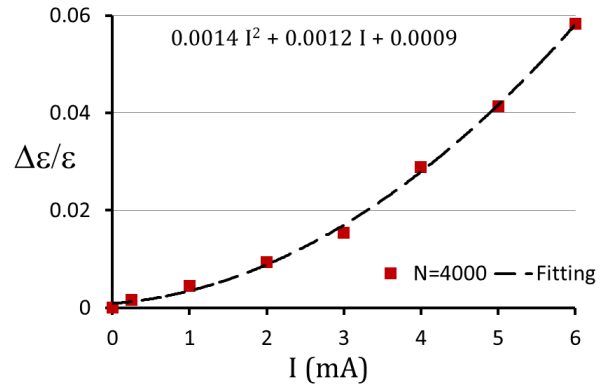}
   \caption{Relative growth of $\epsilon_{6d}$ for varying linac current $I$ (mA), N=4000 and fully isotropic bunch.}
   \label{current}
\end{figure}

We first study the relative increase  in $\epsilon_{6d}$ over 1000 cells for a strictly spherical bunch with Gaussian distribution function, $\epsilon_{x,y}/\epsilon_{z}=1$, $k_{0x,y,z}=60^{o}$ and $k_{x,y,z}\approx 33^{o}$. An example for $N=4000$ is shown in Fig.~\ref{ramp} for an $ rz $ Poisson solver setting with 16x16 grid cells within $3\sigma$. Note that  in TRACEWIN the notion of $n$ cells within $m\sigma$ implies that $ n $ radial or axial grid cells are counted over a radial or axial semi-axis of extent $m\sigma$. Beyond this extent - practically in the beam halo region - a different  method is applied: fields are determined analytically by assuming a Gaussian charge distribution within $m\sigma$. Also note that Fig.~\ref{ramp} indicates a relatively large jitter due to the relatively low number of particles, besides the linear increase of $\epsilon_{6d}$ with distance. 
The explicit dependence of this grid-induced noise on space charge is shown in Fig.~\ref{current}, where the linac current is varied. Here, as in all following graphs, we have introduced a relative growth of $\epsilon_{6d}$ relative to its initial value, i.e. $\Delta\epsilon_{6d}/\epsilon_{6d}$. Throughout this study $\Delta\epsilon_{6d}$ is normalized to 1000 lattice cells, which makes use of a practically linear evolution of $\epsilon_{6d}$ in $s$; except for relatively small values of $N$, where the initial gradient of $\epsilon_{6d}$ was determined and extrapolated to the same value of 1000 cells. We thus obtain from 
Eq.~\ref{modified}: 
\begin{eqnarray}\label{remodified}
\frac{1}{k}\Delta S= \frac{\Delta\epsilon_{6d}}{\epsilon_{6d}}    
=\Delta s \frac{k^{^\star}_f}{3}\:\left(I_A+I_{GN}\right).
\end{eqnarray}

On the current scale of Fig.~\ref{current} the  example of Fig.~\ref{ramp} with a space charge depressed tune of 
$k_{x,y,z}\approx 33^{o}$ is equivalent to $I$=5 mA. 
The effect of space charge results in a steeper than linear, but not quite quadratic dependence on the current (see second order polynomial fitting expression and curve on graph), which needs further study for explanation.

\subsection{Scaling with $N$ and grid in $rz$}\label{sec:Nscaling}
While the emittance exchange in the coherent regime of Fig.~\ref{regimes} changes only slightly for a larger number $N$ of macro particles, the   rates in the noise dominated regime are large for small $N$. Noting that the charge per simulation particle is $\propto 1/N$, we expect a corresponding decrease for large $N$. 

\begin{figure}[h]
   \centering
   \includegraphics*[width=80mm]{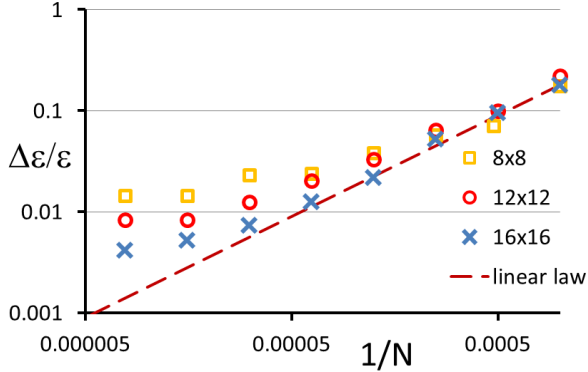}
   \caption{Relative growth of $\epsilon_{6d}$ for $N$ from 1000 to 128.000 macro particles indicating transition to \emph{grid resolution limited regions}.}
   \label{nscaling}
\end{figure}

A more complete picture is shown in  Fig.~\ref{nscaling}, where the relative growth of  $\epsilon_{6d}$, hence the rms entropy, is plotted against $1/N$ as measure of the charge per macro particle (note the double-logarithmic scale, with the dashed line indicating a strictly linear fit through the origin). For the 16x16 grid a linear dependence of $\Delta\epsilon/\epsilon$ is noted in the range from 1000 up to $ \approx $20.000 particles:
\begin{equation}\label{linear}
\Delta\epsilon_{6d}/\epsilon_{6d}\propto N^{-1} 
\end{equation}
 For higher $N$ we find that $\Delta\epsilon/\epsilon$ bends off, however. We also show cases with grid resolution of only 12x12  and find that the departure from a linear law occurs even with less particles, and even more for 8x8 grids. The data indicate that for sufficiently large $ N $ we enter into a \emph{grid resolution limited region}, where  further increasing of $N$ doesn't help much to reduce growth of $\epsilon_{6d}$, unless the grid resolution is increased as well. 
 This transition region is characterized by a typical average number of particles per cell. With the average number of particles per (toroidal) cell in $\Delta r \Delta z$ given as $N/(n_c^{2}*3.14/4) $ we find that the transition occurs typically for 80-100 particles in a toroidal cell (for example for 16.000 particles for $n_c=16$). This suggests that for significantly less particles statistical fluctuations of charges on the grid become large, hence more particles will help to reduce the noise effect. Using significantly more particles, instead, has no further benefit unless the grid resolution is increased. This connection should be considered if one wishes to optimize the efficiency of a simulation with regard to CPU-time. 

\begin{figure}[h]
   \centering
   \includegraphics*[width=85mm]{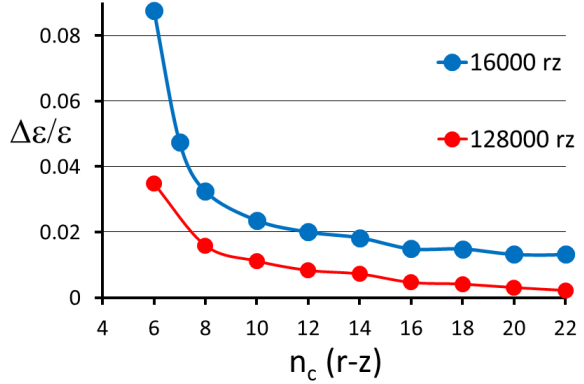}
   \caption{Relative growth of $\epsilon_{6d}$ for $N=16.000/128.000$ and increasing number of grid cells in $r,z$ indicating transition to a \emph{particle number limited region}.}
   \label{cellscaling}
\end{figure}

In Fig.~\ref{cellscaling} we show the rms emittance and entropy growth as function of the number of grid cells $n_{c}$, which is chosen as equal in $r$ and $z$. As observed above, a resolution above 16 cells has no benefit for only 16000 particles. The 128000 particles case, instead, shows progressive benefit for increasing $ n_c $, hence this region is \emph{particle number limited}.  On the other hand, a $ n_c $ as low as 8 shows only a factor 2 reduction of  $\epsilon_{6d}$ by increasing $ N $ from 16000 to 128000 due to the lack of grid resolution.  
In practice, the noise level argument may not be the only one for selecting $N$. Another important consideration is very small fractional beam loss at the level of $ 10^{-4} $ to $ 10^{-6} $. A reliable statistical level can be reached only with an adequately large $N$. 

In summary the observed grid-induced noise appears to be a combined effect of "pseudo-collisions" of particles with a discretised grid and a statistically fluctuating population of the charged particles on this grid.

\section{Temperature anisotropy driven noise}\label{sec:anisotropydriven}
Justification of the separation of $k^{^\star} $ and a "kinetic" term  $I_A+I_{GN}$ needs careful examination. It is not a priori clear that $I_{GN}$ is fully independent of, for example,   space charge, if we notice that the collision distance relative to the focusing period is not independent of space charge. In the following we study the entropy growth as function of the transverse:longitudinal temperature ratio for given N.   
\subsection{Comparison of numerical anisotropy effects with theoretical predictions}\label{sec:num-theor}
In the first example we assume $ k_{0xyz}=60^o$, N=4000, a 16x16 grid with the "6d-ellipsoid" initial distribution. The linac current is chosen again such that $ k_{xyz}=33^o$ for equal emittances. Different emittances in transverse and longitudinal directions are realized in such a way that the product $ \epsilon_x\epsilon_y \epsilon_z$ remains invariant. The initial temperature anisotropy on the abscissa of Fig.~\ref{symaniso} is determined via Eq.~\ref{ratio} using actual emittances and space charge dependent tunes $k$ from TRACEWIN. In Fig.~\ref{symaniso} we show  $ \Delta\epsilon_{6d}/\epsilon_{6d} $ determined from the initial gradient of the 6d emittance from TRACEWIN data (taken over the first 100 or few hundred cells):
\begin{figure}[h]
   \centering
   \includegraphics*[width=85mm]{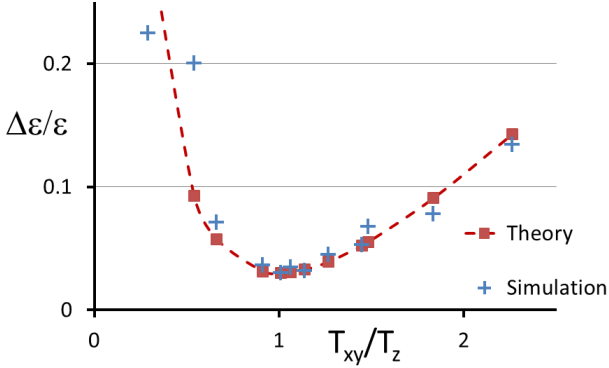}
   \caption{Relative growth of $\epsilon_{6d}$ for $ k_{0xyz}=60^o$ and varying (initial) temperature anisotropy.}
   \label{symaniso}
\end{figure}

The simulation results  confirm the existence of a minimum emittance growth of $\Delta\epsilon_{6d}/\epsilon_{6d}\approx 0.03$ at the point, where the temperature ratios $r$ in Eq.\ref{anisotropies} are all equal to unity. In order to check the applicability of the analytical $ I_A $ in Eq.~\ref{emit-evol} we need to fit $ k_f $ and $I_{GN}$ to the numerical data using Eq.\ref{remodified}.
The thus resulting "theoretical" curve in Fig.~\ref{symaniso} shows good agreement with the simulation data. 

\begin{figure}[h]
   \centering
   \includegraphics*[width=85mm]{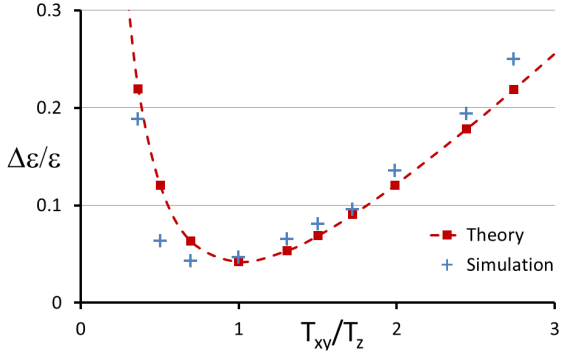}
   \caption{ $\Delta\epsilon_{6d}/\epsilon_{6d}$ for $ k_{0xyz}=60/60/47^o$ and varying (initial) temperature anisotropy.}
   \label{asymaniso}
\end{figure}

We have also tested this comparison for split $ k_{0xy} $ and $k_{0z} $ to $ 60/60/47^o $ - this time using a Gaussian input distribution - and find reasonably good agreement as shown in Fig.~\ref{asymaniso}. The isotropic offset is slightly higher - 0.04 -, which might be connected with the stronger tune depression in the weaker focusing direction ($ z $).
\subsection{Scaling with N}\label{sec:nscaling}
\begin{figure}[h]
   \centering
   \includegraphics*[width=85mm]{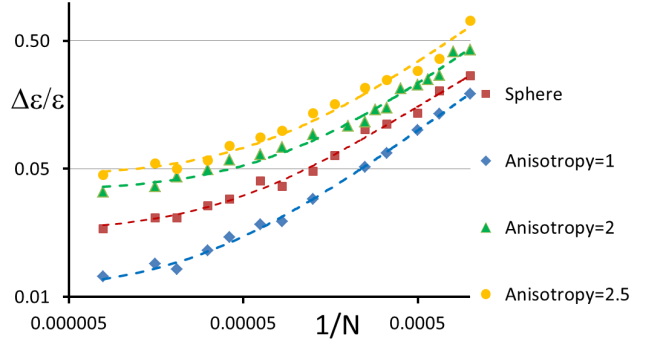}
   \caption{ $\Delta\epsilon_{6d}/\epsilon_{6d}$ for different anisotropic cases and $ k_{0xyz}=60/60/47^o$ as function of $N$ (with spherical bunch and $k_{0xyz}=60^o$ for comparison).}
   \label{asymanisoN}
\end{figure}

The separation of $N-$dependence and of anisotropy for the same focusing and initial distribution is further examined in the double-logarithmic graphs of Fig.~\ref{asymanisoN}. For large $N$ (above $\approx 10^4$ in our examples) we retrieve the typical behaviour of deviation from linear scaling in $N^{-1}$ already observed in Fig.~\ref{nscaling} and interpreted as effect of insufficient lattice resolution in spite of sufficient particle statistics. 
It is noted that in this linear scaling regime the curves for different anisotropy are basically shifted by constants, which supports that $ N $ doesn't enter into $ I_{A} $, but only into $ k_{f} $.
\section{3d bunches with $xyz$ Poisson solver.}\label{sec:numerical-xyz}

The findings obtained for $rz-$symmetry have also an equivalent in the use of the $xyz$ Poisson solver option. The $n_{cx,y,z}$ are understood here as half number of cells between the maximum grid extent values.  We also consider variation of the number of space charge  "steps of calculation" - keeping the number of lattice calculations fixed at 20/m.

\subsection{XYZ versus RZ Poisson solver}\label{sec:xyz-rz}
\begin{figure}[h]
   \centering
   \includegraphics*[width=85mm]{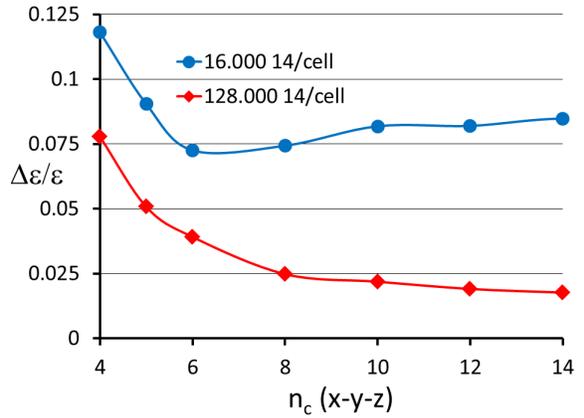}
   \caption{Relative growth of $\epsilon_{6d}$ for $N$=16.000/128.000, as function of number of grid cells in $x,y,z$ for periodic solenoid lattice of Fig.~\ref{env}.}
   \label{xyzscaling}
\end{figure}

We first consider the identical periodic solenoidal case used in Figs.~\ref{env},~\ref{nscaling},~\ref{cellscaling} and replace the $ rz $ Poisson solver by  the $ xyz $ option. This is shown in Fig.~\ref{xyzscaling} employing (as before) 14 space charge steps per lattice cell. However, we find that for 16.000 particles  $\frac{\Delta \epsilon}{\epsilon}$saturates around 0.08 - mor than 5 times larger than for the $rz$ solver.  For 128.000 particles, instead,             the $ xyz$-solver results are roughly only a factor of 2  larger than  the corresponding values for the $ rz$-solver.

For $ N$=16.000 we note that increasing the number of space charge kicks from seven to eleven per cell leads to only half the growth of $\epsilon_{6d}$ for values of $ n_c \approx 10$. Less frequent space charge kicks are "harder", and it is understandable that they enhance entropy growth. This difference, however, is less pronounced for larger $N$.

The observation of "particle statistics limited" and "grid resolution limited" is even more obvious than for the $ rz $ Poisson solver:

\begin{itemize}
\item \emph{Grid resolution limited:} For small $ n_c $ there is only weak benefit by increasing  $N$ from 16.000 to 128.000 due to insufficient grid resolution 
\item  \emph{Particle statistics limited:} Increasing $n_{c}$ above 6 enhances again the emittance growth for the case $ N$=16.000; in this case the average number of particles per grid cell falls under 10 and statistics appears insufficient. Raising $ N $ to 128.000 has visible benefit and no saturation is seen, although the improvement above $n_{c}\approx 8...10$ is only minor.
 \item \emph{Optimum grid resolution:} There is evidence that for each $N$ an optimum grid resolution exists, where the emittance growth is smallest; the corresponding $n_{c}$ appears to be the higher the larger $N$. 
\end{itemize}
\subsection{FODO lattice}\label{sec:fodo}
\begin{figure}[h]
   \centering
   \includegraphics*[width=85mm]{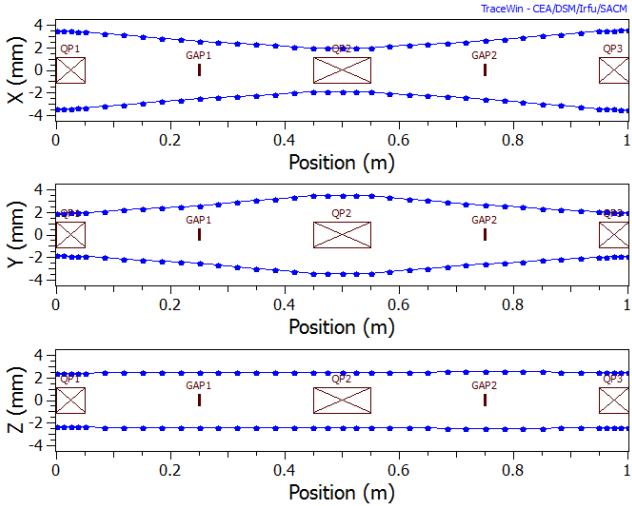}
   \caption{Equivalent FODO lattice with rf gaps.}
   \label{fodo}
\end{figure}

\begin{figure}[h]
   \centering
   \includegraphics*[width=85mm]{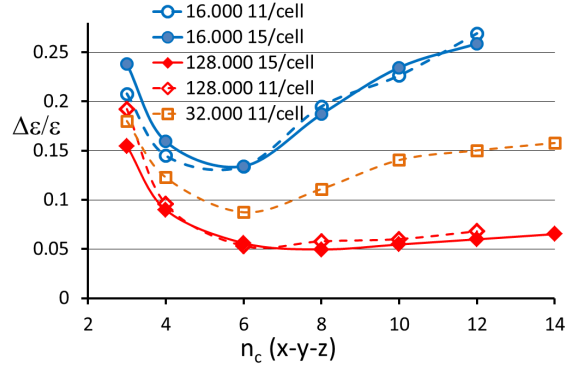}
   \caption{Relative growth of $\epsilon_{6d}$ in FODO for $N$=16.000/32.000/128.000, as function of the number of grid cells in $x,y,z$ and different numbers of space charge steps.}
   \label{fodoscaling}
\end{figure}

We adopt an "equivalent" periodic FODO lattice with rf gaps and   cell length again 1 m as shown in Fig.~\ref{fodo}. The same phase advance per m is chosen as in the previous examples as well as identical emittances, which results in approximately the same space charge depressed tunes. 
The alternating focusing causes a strong modulation and local imbalance of "temperatures", which is described in Ref.~\cite{struck1996} as source of entropy growth. We expect that this mechanism amplifies respectively adds to the already mentioned "non-Markov" effects found for the periodic solenoid. This is shown in Fig.~\ref{fodoscaling} as function of $n_c$ and for different $N$ as well as space charge steps. 
It is noted that the difference by decreasing the number of space charge steps per cell from 15 (10/m in TRACEWIN) to 11 (3/m in TRACEWIN) is minor.  Both cases, low and high particle numbers, show again that an optimum $ n_c $ exists, which is the higher the larger $ N $. As before, large $N$ is efficient only if the grid resolution is  sufficiently large.

Comparing with Fig.~\ref{xyzscaling} for the periodic solenoid we find an enhanced noise level. The results of Fig.~\ref{fodoscaling} also differs from the result reported in Ref.~\cite{pichoff1998}, where low and high particle number (transverse) emittance growth is nearly identical around $ n_c=8 $.  For this grid resolution we find that 16.000 particles lead to practically  four times the growth in $\epsilon_{6d}$ than 128.000 particles. Theoretically, following the result of Fig.~\ref{nscaling} and Eq.~\ref{linear},  the difference could be even as large as a factor eight. In practice, such high resolution is not feasible - in particular jointly with large $ N $. 
\subsection{DTL lattice}\label{sec:dtl}
\begin{figure}[h]
   \centering
   \includegraphics*[width=85mm]{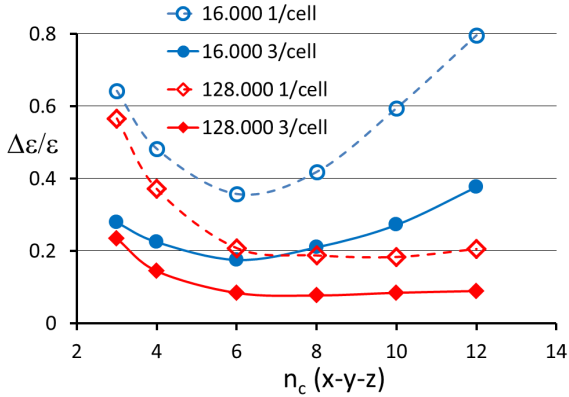}
   \caption{DRIFT TUBE LINAC CELL option for FODO: Relative growth of $\epsilon_{6d}$ for $N$=16.000/128.000, as function of number of grid cells in $x,y,z$ in same lattice structure as Fig.~\ref{fodoscaling}.}
   \label{fododtlcell}
\end{figure}

In TRACEWIN - when applied to linac design - it is customary to employ the compressed and faster option of DRIFT TUBE LINAC CELLS. It simplifies space charge calculation to one single step of calculation per DTL cell, or optionally to three steps (marked by 1/cell and 3/cell). We apply this method to the cell shown in Fig.~\ref{fodo} and obtain the results shown in Fig.~\ref{fododtlcell}
The results indicate that the frequently used 1/cell option implies a not insignificant additional noise level. 
\section{Conclusion and Outlook}
We have explored space charge and grid induced noise in 3d PIC simulation with the TRACEWIN code, and compared results with the analytical rms entropy growth equation by Struckmeier based on a collisional approach. Our findings confirm that the 6d rms emittance - defined as product of the three 2d rms emittances - is a suitable measure for the numerical noise and entropy growth generated by space charge and grid effects.  It can thus be used as "quality measure" for optimizing simulation of high intensity beams. In particular, we have found that an optimum combination of $N$ and grid resolution exists: further increasing of the grid  resolution is useful only, if the simulation particle number (per grid cell) is increased as well. Likewise, a larger number of simulation particles has little benefit unless there is sufficient grid resolution.

We also find that our simulations are in a grid effect dominated regime, which differs from the assumption of a  collisional regime assumed in the work by Struckmeier. Thus, we obtain entropy growth even in fully isotropic cases with no temperature differences. Further work is needed to explore in more detail the transition between two distinct regimes: the "grid effect dominated regime" - claimed here - where particles interact in a non-Liouvillean way with the  charge distribution on a grid; and a "particle collision regime", where (Markov type) particle-particle collisions are resolved. The latter can be assumed to be increasingly relevant at much higher grid resolution than employed here. In the grid dominated  regime case it should also matter that our time step is not sufficiently small compared with a typical transit time across a grid cell. Also, the presence of "coherent flow" in periodic focusing - as opposed to incoherent temperature-like motion typical for steady state plasmas - must in a suitable way enter into a future more detailed analysis.

 Further work is required to extend this analysis to 3d bunches with substantially different oscillation frequencies - like slow synchrotron motion - in order to explore PIC simulation noise for circular accelerators in a similar framework.
 
 {\bf Acknowledgment:}
 The authors are grateful for valuable discussions with J. Struckmeier.

\end{document}